# Analysis of Measurements of the Magnetic Flux Density in Steel Blocks of the Compact Muon Solenoid Magnet Yoke with Solenoid Coil Fast Discharges

Vyacheslav Klyukhin [1,2,3,]*, Benoit Curé [2], Andrea Gaddi [2], Antoine Kehrli [2], Maciej Ostrega [2] and Xavier Pons [2]

1. Skobeltsyn Institute of Nuclear Physics, Lomonosov Moscow State University, RU-119991 Moscow, Russia
2. CERN, CH-1211 Geneva 23, Switzerland; benoit.cure@cern.ch (B.C.); andrea.gaddi@cern.ch (A.G.); antoine.kehrli@cern.ch (A.K.); maciej.ostrega@cern.ch (M.O.); xavier.pons@cern.ch (X.P.)
3. Joint Institute for Nuclear Research, RU-141980 Dubna, Russia
* Correspondence: vyacheslav.klyukhin@cern.ch

**Abstract:** The general-purpose Compact Muon Solenoid (CMS) detector at the Large Hadron Collider (LHC) at CERN is used to study the production of new particles in proton–proton collisions at an LHC center of mass energy of 13.6 TeV. The detector includes a magnet based on a 6 m diameter superconducting solenoid coil operating at a current of 18.164 kA. This current creates a central magnetic flux density of 3.8 T that allows for the high-precision measurement of the momenta of the produced charged particles using tracking and muon subdetectors. The CMS magnet contains a 10,000 ton flux-return yoke of dodecagonal shape made from the assembly of construction steel blocks distributed in several layers. These steel blocks are magnetized, with the solenoid returned magnetic flux and wrap the muons escaping the hadronic calorimeters of total absorption. To reconstruct the muon trajectories, and thus to measure the muon momenta, the drift tube and cathode strip chambers are located between the layers of the steel blocks. To describe the distribution of the magnetic flux in the magnet yoke layers, a three-dimensional computer model of the CMS magnet is used. To validate the calculations, special measurements are performed, with the flux loops wound in 22 cross-sections of the flux-return yoke blocks. The measured voltages induced in the flux loops during the CMS magnet ramp-ups and -downs, as well as during the superconducting coil fast discharges, are integrated over time to obtain the initial magnetic flux densities in the flux loop cross-sections. The measurements obtained during the seven standard ramp-downs of the magnet were analyzed in 2018. From that time, three fast discharges occurred during the standard ramp-downs of the magnet. This allows us to single out the contributions of the eddy currents, induced in steel, to the flux loop voltages registered during the fast discharges of the coil. Accounting for these contributions to the flux loop measurements during intentionally triggered fast discharges in 2006 allows us to perform the validation of the CMS magnet computer model with better precision. The technique for the flux loop measurements and the obtained results are presented and discussed. The method for measuring magnetic flux density in steel blocks described in this study is innovative. The experience of 3D modeling and measuring the magnetic field in steel blocks of the magnet yoke, as part of a muon detector system, has good prospects for use in the construction and operation of particle detectors for the Future Circular Electron–Positron Collider and the Circular Electron–Positron Collider.

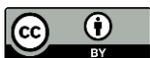

## 1. Introduction

The Compact Muon Solenoid (CMS) multi-purpose detector [1] at the Large Hadron Collider (LHC) [2] registers the charged and neutral particles created in proton–proton collisions at a center of mass energy of 13.6 TeV. The detector includes a 6 m diameter superconducting solenoid coil [3] with a length of 12.5 m and a central magnetic flux density $B_0$ of 3.81 T that is created by the operational direct current of 18.164 kA. The major particle subdetectors are located inside the superconducting coil and around the interaction point of the proton beams: the silicon pixel and strip tracking detectors register the charged particles; the solid crystal electromagnetic calorimeter registers electrons, positrons and gamma rays; and barrel and endcap hadronic calorimeters of total absorption register the energy of all the hadron particles. The coil is inserted into a 10,000-ton flux-return steel yoke that consists of five three-layered dodecagonal barrel wheels positioned around the coil, and several endcap disks at both ends of the coil. The drift tube chambers [4] are installed between the layers of the barrel wheels and serve to measure the momenta of muons escaping the barrel calorimeters. The muon trajectories are bent in the steel blocks of the barrel layers magnetized with the solenoid returned magnetic flux. This bending allows for muon momentum measurements with drift tube chambers, and knowledge of the magnetic flux density distribution in the steel blocks of the yoke is key to the measurement precision. In a similar manner, the cathode strip chambers [5] located between the magnetized endcap disks measure the muon momenta in the detector forward region.

The magnetic flux density inside the solenoid coil has been measured with a special field-mapping machine to a relative precision of $7 \times 10^{-4}$. Inside the coil, the magnetic flux density is monitored by six nuclear magnetic resonance probes with an averaged precision of $(5.2 \pm 1.3) \times 10^{-5}$ T [6]. In the air gaps between the yoke barrel wheels and on the endcap disks' surfaces, the magnetic flux density is monitored with three-dimensional (3D) B-sensors (Hall probes) with an averaged precision of $(3.5 \pm 0.5) \times 10^{-5}$ T [6]. The magnetic flux in the steel blocks of the magnet flux-return yoke has been calculated using a CMS magnet 3D model [7] based on the program TOSCA (TwO SCAlar potential method) [8], developed in 1979 [9] at the Rutherford Appleton Laboratory. This model reproduces the magnetic flux density distribution measured with the field-mapping machine inside the CMS coil to within 0.1% accuracy [6]. The CMS magnetic field map [10] prepared with this model inside a cylindrical volume of 9 m radius includes the steel yoke layers and has a rotational symmetry.

To verify the magnetic flux distribution calculated in the yoke steel blocks, direct measurements of the magnetic flux density in the selected regions of the yoke were performed during the CMS magnet test in 2006 [11], when four fast discharges of the CMS coil (190 s time constant) from the currents of 12.5, 15, 17.55, and 19.14 kA were triggered manually to test the magnet protection system. These discharges were used to induce voltages with amplitudes of 0.5–4.5 V in the 22 flux loops wound around 12 yoke blocks in special grooves 30 mm wide and 12–13 mm deep. The loops have 7 to 10 turns of a 45-strand flat ribbon cable. The cross-sections of areas enclosed by the flux loops vary from 0.3 to 1.59 m² on 10 blocks of the yoke barrel wheels and from 0.5 to 1.12 m² on 2 blocks of the yoke endcap disks [12]. An integration technique [13] was developed to reconstruct the average initial magnetic flux density in the cross-sections of these selected steel blocks at full magnet excitation.

At that time, no fast discharge of the CMS magnet from its operational current of 18.164 kA, which corresponds to a central magnetic flux density of 3.81 T, was performed. To measure the magnetic flux density in the steel blocks of the flux-return yoke at this operational current, seven standard linear discharges of the CMS magnet with a current rate as low as 1–1.5 A/s were performed later [6,14]. To provide these measurements, the voltages induced in the flux loops (with amplitudes of 20–250 mV) were recorded with six 16-bit data acquisition (DAQ) modules and integrated offline over time [6].

In 2022, these modules were replaced with a new 16-bit DAQ system based on the Siemens S7-1500 programmable logic controller (PLC) [15]. The new readout boxes were connected to each other by Ethernet cables through the CMS cable chains, which allow the magnet yoke to open or close without disconnection of the DAQ system parts. This new organization of the readout scheme allowed us to register two fast discharges of the CMS coil which occurred unintentionally: one from the operational current of 18.164 kA, and another one from the current of 15.221 kA that happened during the magnet standard discharge from the operational current to this value. Both these fast discharges, in combination with the fast discharge from 9.5 kA registered in 2017 [14], are used in this study to single out the contribution of the eddy currents into the induced voltages registered in the fast discharges of 2006 and to improve the accuracy of these flux loop measurements.

The article is organized as follows: Section 2 describes the apparatus used for the flux loop measurements and the procedure for estimating the eddy current contribution; Section 3 contains the results of the flux loop measurements in the fast discharges of the CMS coil; Section 4 presents the discussion of the obtained results; and finally, conclusions are drawn in Section 5.

## 2. Materials and Methods

### 2.1. Description of the Flux Loop Measuring System

In Figure 1, an area of the flux loop location in the CMS magnet yoke is shown [6]. The magnetic flux density distribution in this area is described with a color scale from zero to 4 T with a unit of 0.5 T. The positions of the flux loop cross-sections are indicated by the black lines across the magnetized steel blocks. Sixteen flux loops located on ten barrel yoke steel blocks in the 30° azimuthal sector at 270° measure the negative values of the axial magnetic flux density $B_z$ that is orthogonal to the flux loop cross-sections. Six flux loops installed on the 18° azimuthal sector of the endcap disks at 270° measure the positive values of the vertical magnetic flux density $B_y$ that is also orthogonal to the flux loop cross-sections. The Y- and Z-positions of the 3D B-sensors used to monitor the magnetic flux density in the air gaps between the steel blocks are indicated in Figure 1 by small black squares.

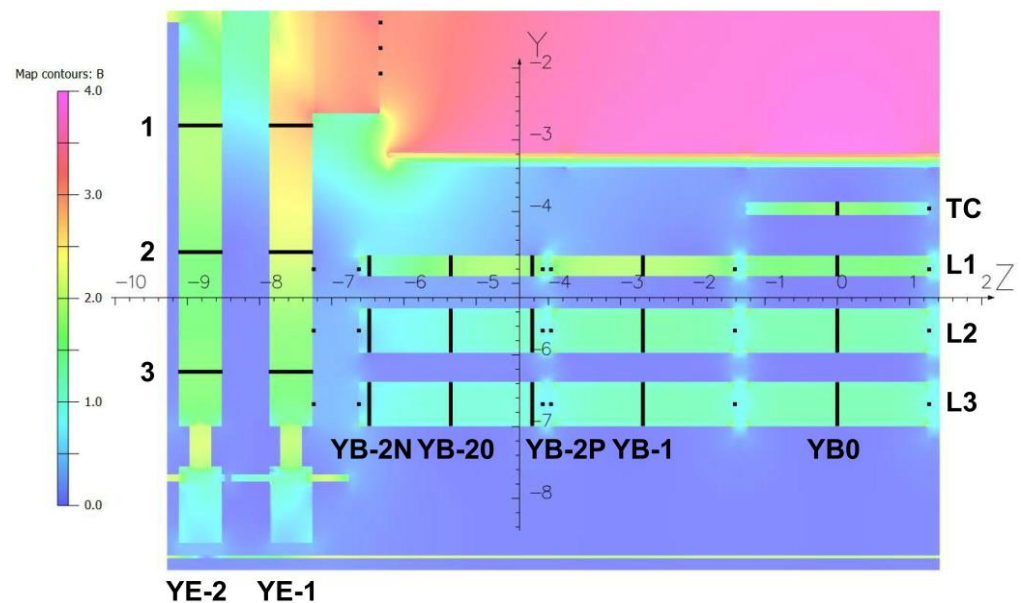

**Figure 1.** Modeled distribution of the magnetic flux density $B$ in Tesla in the vertical YZ-plane of the flux loop location area. Sixteen flux loops are installed in the 30° azimuthal sector at 270° of the CMS magnet barrel flux-return yoke on four layers (TC, L1, L2, L3) of the central barrel wheel YB0, and on three layers (L1, L2, L3) of the barrel wheels YB−1 and YB−2 at negative Z-coordinates, shown in meters on the Z-axis. Six flux loops are wound at the distances 1, 2, and 3 along the 18° azimuthal sector at 270° of the CMS endcap thick disks YE−1 and YE−2 [6]. The black lines indicate the Y- or Z-positions of the loop cross-sections. The small black squares indicate the projections to the YZ-plane of the 3D B-sensor locations installed on the surfaces of the yoke steel blocks.

The coordinate axes shown in Figure 1 represent the CMS coordinate system, where the origin of the CMS reference frame is located in the center of the superconducting solenoid coil, the *X* axis lies in the LHC plane and is directed to the center of the LHC machine, the *Y* axis is directed upward and is perpendicular to the LHC plane, and the *Z* axis makes up the right triplet with the *X* and *Y* axes and is directed along the vector of magnetic flux density created on the axis of the superconducting coil. The direction of the axial magnetic flux density inside (red area in Figure 1) and outside (blue area in Figure 1) the coil is opposite.

To measure the magnetic flux density induced in the steel blocks with the coil-returned magnetic flux, a variation in the magnetic flux though the flux loop cross-sections is needed. This variation creates an analog voltage in the strands of each flux loop, and this voltage can be described as a function of time *t* by the following equation:

$$v(t) = \frac{d\Phi}{dt} = A \cdot N \cdot \frac{dB_i}{dI} \cdot \frac{dI}{dt}. \qquad (1)$$

Here, *Φ* is the magnetic flux through the flux loop cross-section, *A* is the area encircled by the flux loop, *N* is the number of turns in the flux loop, $B_i$ is either the axial $B_z$ or the vertical $B_y$ magnetic flux density component, and *I* is the current in the solenoid coil. To obtain the magnetic flux at the beginning of the magnetic field variation, the voltage *v(t)* should be integrated over the time of the variation [13]. Dividing the integrated magnetic flux *Φ* by the flux loop area *A* and the number of the loop turns *N* gives the initial magnetic flux density $B_i$.

To induce the voltages in the flux loops, either the standard magnet ramp-downs [14] or the coil current fast discharges [11] can be used. The present analysis is based on seven fast discharges, as shown in Figure 2 and performed from the magnet currents of 9.5, 12.5, 15, 15.221, 17.55, 18.164, and 19.14 kA. These currents create initial central magnetic flux densities of $B_0$ of 2.02, 2.64, 3.16, 3.20, 3.68, 3.81, and 4.01 T, respectively, in the CMS superconducting coil.

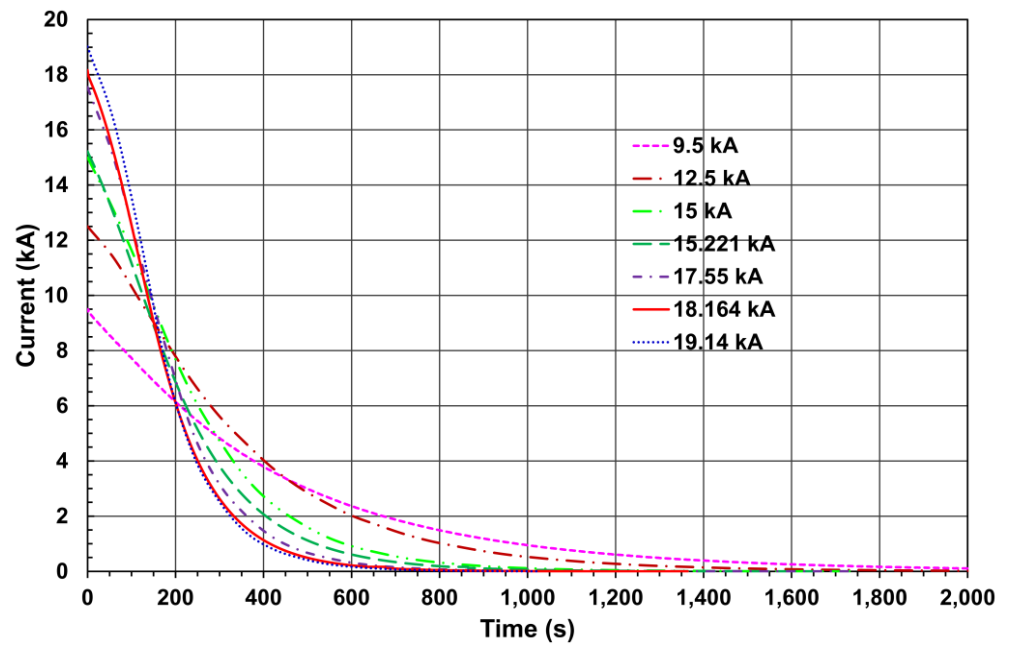

**Figure 2.** Measured magnetic current variations during the fast discharges which occurred from the CMS magnet currents of 9.5, 12.5, 15, 15.221, 17.55, 18.164, and 19.14 kA. These currents create initial central magnetic flux densities of $B_0$ of 2.02, 2.64, 3.16, 3.20, 3.68, 3.81, and 4.01 T, respectively, in the CMS superconducting coil.

The voltages induced in the flux loops have been registered with three different DAQ systems. The first one [12] was used to register the flux loop voltages in the fast discharges performed in 2006 [11] from the initial CMS magnet currents of 12.5, 15, 17.55, and 19.14 kA. This system included seven USB-based DAQ modules USB-1208LS (Measurement Computing Corporation: Norton, MA, USA) with four differential 12-bit analog inputs each. The USB-1208LS DAQ modules were attached by the USB cables to two network-enabled AnywhereUSB®/5 hubs (Digi International Inc.: Minnetonka, MN, USA) connected to a personal computer through a 3Com® OfficeConnect® Dual Speed Switch 5 (3Com Corporation: Marlborough, MA, USA) sitting on a local Ethernet network cable of 90 m, which connected the switch to the computer containing the TracerDAQ® Pro data acquisition software from Measurement Computing. The precision of the signal amplitude measurements with the 12-bit modules was 2.44 mV.

In 2013/2014, this DAQ system was upgraded by replacing the 12-bit modules with new 16-bit USB-1608G modules from the same manufacturer. This replacement has allowed the precision of the signal amplitude measurements to be improved to 0.15 mV [14]. The new 16-bit readout gives a resolution of 0.75% at a typical signal amplitude of 20 mV. The local Ethernet network cable of 90 m has been replaced by the shielded optical fiber cable of 100 m, supplied with two Magnum CS14H-12VDC hardened Convertor Switches, from GarrettCom, Inc, on both ends. This modification has allowed the flux loop measurements to be performed during the CMS magnet standard ramp-downs, with a current discharge speed as low as 1–1.5 A/s at acceptable accuracy [14], with 20 to 250 mV measured voltage amplitudes. With this system, the flux loop voltages induced during the CMS magnet fast discharge from the current of 9.5 kA were also recorded in 2017 [6,14].

The present DAQ scheme, developed in 2022 and used to register the flux loop voltages during the CMS magnet fast discharges from the currents of 15.221 and 18.164 kA, is displayed in Figure 3.

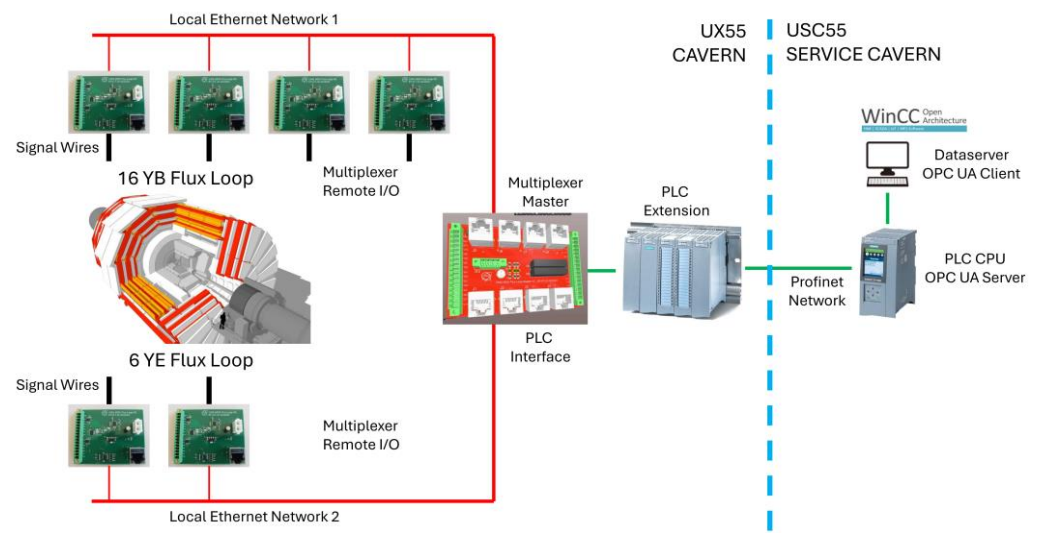

**Figure 3.** DAQ block diagram of the flux loop signal processing. The measuring system contains six custom multiplexers assembled in the patch boxes to read out the voltages induced in the flux loops. Through the custom-made master PLC interface circuit, the multiplexers communicate to Siemens S7-1500 PLC drives to propagate the signals to the WinCC OA software that stores the voltage values in the database.

This DAQ design includes six radiation-hardened multiplexers assembled with 24 V DC power supplies and Burndy connectors in the custom patch boxes shown in Figure 4a.

Three, four, or six flux loops are wired to each multiplexer through the 19-pin Burndy connector shown in Figure 4b. The multiplexer interface circuit allows up to eight channels to be read through the Ethernet cable. All the Ethernet cables from the multiplexers are connected to a main circuit interface, which dispatches the analog signals to the PLC modules and the control bits of the multiplexers to a digital output module. The PLC modules are connected to the Siemens S7-1500 CPU [15] that contains the readout software and is located outside of the experimental cavern. The information is sent by the PLC through the WinCC OA software v. 3.19 and OPC UA driver to the database to be archived.

The main feature of the new DAQ design is a waiting regime in the signal readout. The signals from the flux loops are read with a rate of 1 Hz, sending them channel-by-channel to the database. This procedure allows for the registration of the flux loop signals at the unpredictable fast discharges which occurred during magnet operation.

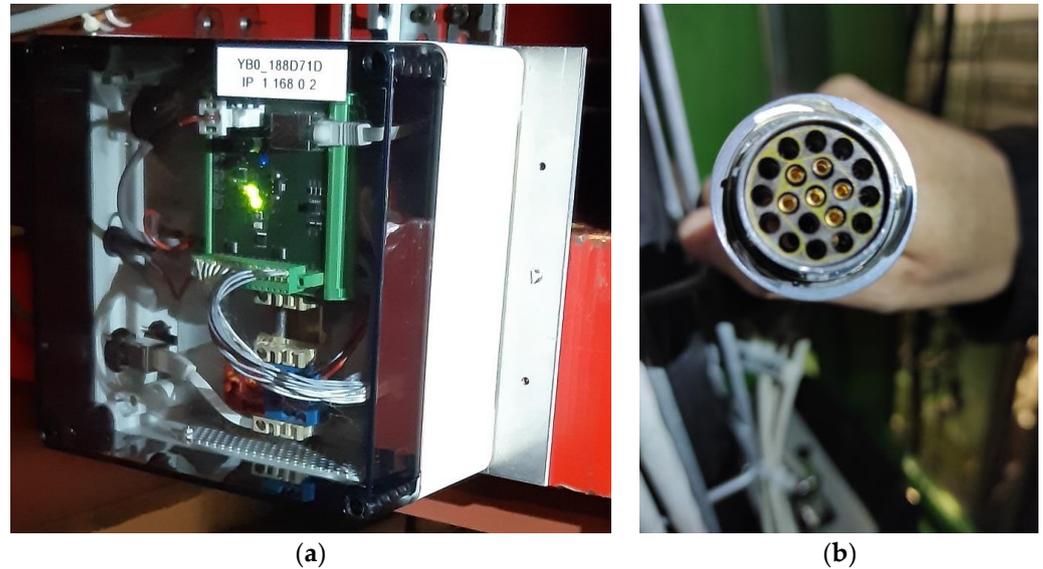

(**a**) (**b**)

**Figure 4.** (**a**) One of six readout boxes with a custom multiplexer, a 24 V DC power supply, and a 19-pin Burndy connector (**b**) wiring from three to six flux loops to each multiplexer.

The fast discharges induce voltages with rather complicated time dependence in the flux loops. Examples of these time dependence shapes in the flux loops YB−2P/L2 and YE−2/2 during different fast discharges are shown in Figure 5.

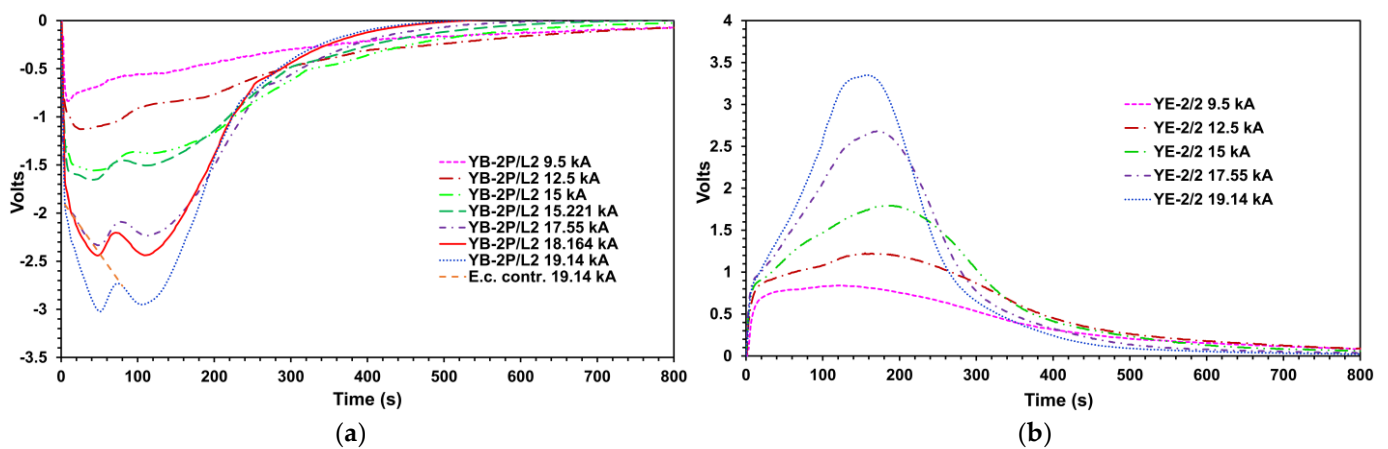

(**a**) (**b**)

**Figure 5.** (**a**) Voltages measured in the flux loop P on the L2 layer of the external barrel wheel YB−2 during all seven fast discharges performed with different initial magnet currents. The orange short-dashed line cuts the contribution of the eddy currents to the signal at the beginning of the magnet fast discharge performed from the current of 19.14 kA. This contribution has been estimated [6] at the level of 4.6% into the integrated magnetic flux. (**b**) Voltages measured in the middle flux loop 2 of the endcap disk YE−2 during five fast discharges performed from different magnet currents.

The voltage shapes of each curve shown in Figure 5a have two minima and one extreme between them. The second minima occurred at 116–185 s from the beginning of the fast discharges and corresponded to the inflection points of the curves of the current dependencies on time shown in Figure 2. The first minima occurred at 9.25–51 s from the beginning of the fast discharges and corresponded to the maxima of the eddy currents induced in the steel cross-section of the flux loop YB−2P/L2 by changing the current rate. The eddy currents are damped due to the resistivity of steel near the extrema, which occurred at 71–122 s. The area of the signal from the first minimum to the extremum describes the eddy current contribution to the induced voltage and decreases relatively with increasing initial currents of the fast discharges. The area separated by the orange short-dashed line from the signal at the fast discharge from the current of 19.14 kA corresponds to the contribution of the eddy current into the integrated voltage at the level of 4.6% [6].

*2.2. Estimation of the Eddy Current Contribution into the Induced Voltages*

To estimate the eddy current contribution more precisely, the three fast discharges which occurred occasionally during the standard ramp-downs of the CMS magnet from the operational current of 18.164 kA were used. These three fast discharges are presented in Figure 6 and have been generated from the currents of 9.5 (on 30 November 2017), 15.221 (on 8 August 2023), and 18.164 kA (on 22 March 2023) in several time intervals from the beginning of the standard ramp-downs from the current of 18.164 kA. The measured voltages induced in each flux loop according to Equation (1) have been integrated over the entire time of the magnet discharges, i.e., during 17,000 (9.5 kA), 4460 (15.221 kA), and 1363 (18.164 kA) s. All these integrations give the initial magnetic fluxes $\Phi_{FD}$ in each flux loop cross-section. To estimate the eddy current contributions, these initial magnetic fluxes are compared with the reference magnetic fluxes $\Phi_{SRD}$ obtained by averaging the fluxes in each flux loop obtained in seven measurements performed with the standard magnet ramp-downs [6] from the current of 18.164 kA.

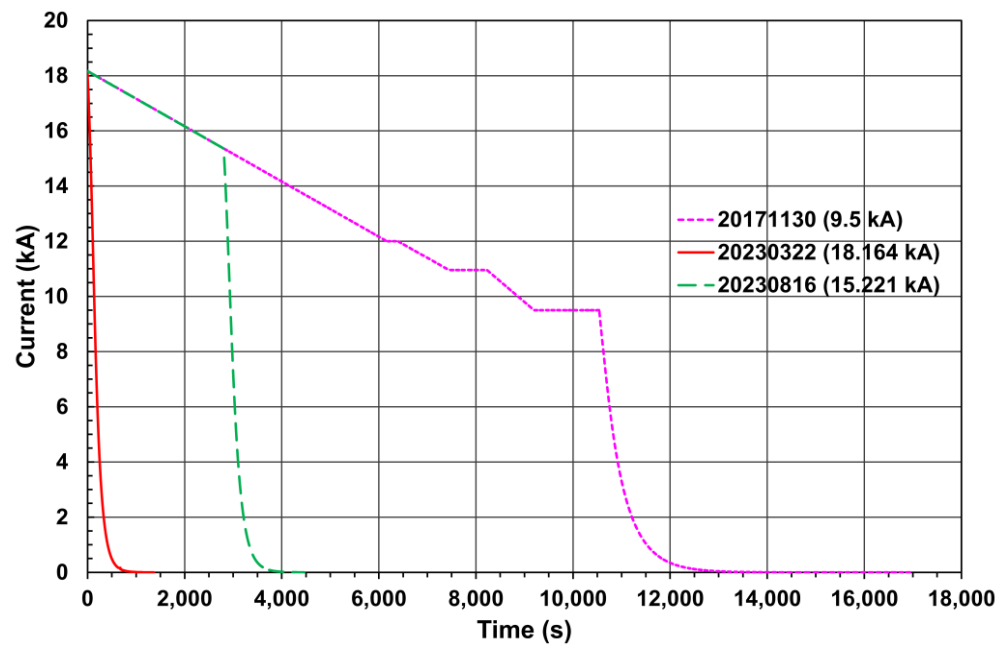

**Figure 6.** Three fast discharges occasionally occurred from the currents of 9.5 (on 30 November 2017), 15.221 (on 8 August 2023), and 18.164 (on 22 March 2023) kA during the standard CMS magnet ramp-downs with a rate of 1–1.5 A/s from the operational magnet current of 18.164 kA.

The estimated contributions of the eddy current, *E.c. contr*, in each flux loop are derived from a ratio:

$$E.c.\ contr = \frac{(\Phi_{FD} - \Phi_{SRD})}{\Phi_{SRD}} \cdot 100\%. \tag{2}$$

Here, $\Phi_{FD}$ is the magnetic flux through the given flux loop measured with the standard ramp-down from the current of 18.164 kA and the subsequent fast discharge of the magnet, and $\Phi_{SRD}$ is the magnetic flux through the given flux loop averaged over seven measurements performed with the standard magnet ramp-downs from the same current of 18.164 kA.

In the barrel wheels, the averaged magnetic fluxes, $\Phi_{SRD}$, through the cross-sections of the flux loops in the standard magnet ramp-downs have a minimum of 120.87 Wb (in the flux loop YB−2N/L1) and a maximum of 792.23 Wb (in the flux loop YB0/L3). In the endcap disks, the similar magnetic fluxes have a minimum of 438.49 Wb (in the flux loop YE−2/1) and a maximum of 833.96 Wb (in the flux loop YE−2/3).

In the barrel wheels, the magnetic fluxes, $\Phi_{FD}$, integrated in each flux loop during the standard ramp-down from the current of 18.164 kA followed by the fast discharge from the current of 9.5 kA (on 30 November 2017) have a minimum of 125.53 Wb (in the flux loop YB−2N/L1) and a maximum of 814.02 Wb (in the flux loop YB0/L3). In the endcap disks, the similar magnetic fluxes have a minimum of 471.37 Wb (in the flux loop YE−2/1) and a maximum of 837.26 Wb (in the flux loop YE−2/3).

With these comparisons, Equation (2) gives the contributions of the eddy currents, *E.c. contr*, as follows: 3.86% in the voltage induced in the flux loop YB−2N/L1, 2.75% in the voltage induced in the flux loop YB0/L3, 7.50% in the voltage induced in the flux loop YE−2/1, and 0.40% in the voltage induced in the flux loop YE−2/3 by the fast discharge from the current of 9.5 kA.

The values of the magnetic fluxes $\Phi_{FD}$ obtained with the magnet discharges of 8 August 2023 and 22 March 2023, are below the values obtained in the discharge of 30 November 2017, and above the averaged magnetic fluxes $\Phi_{SRD}$.

The contributions of the eddy currents for the other four initial currents of the fast discharges are obtained by the interpolation and extrapolation of the *E.c. contr* values from three reference discharges of 30 November 2017, 8 August 2023, and 22 March 2023, using the polynomial of the second order.

The results of comparisons for 20 flux loops and for all seven fast discharges from the currents of 9.5, 12.5, 15, 15.221, 17.55, 18.164, and 19.14 kA are presented in Figure 7.

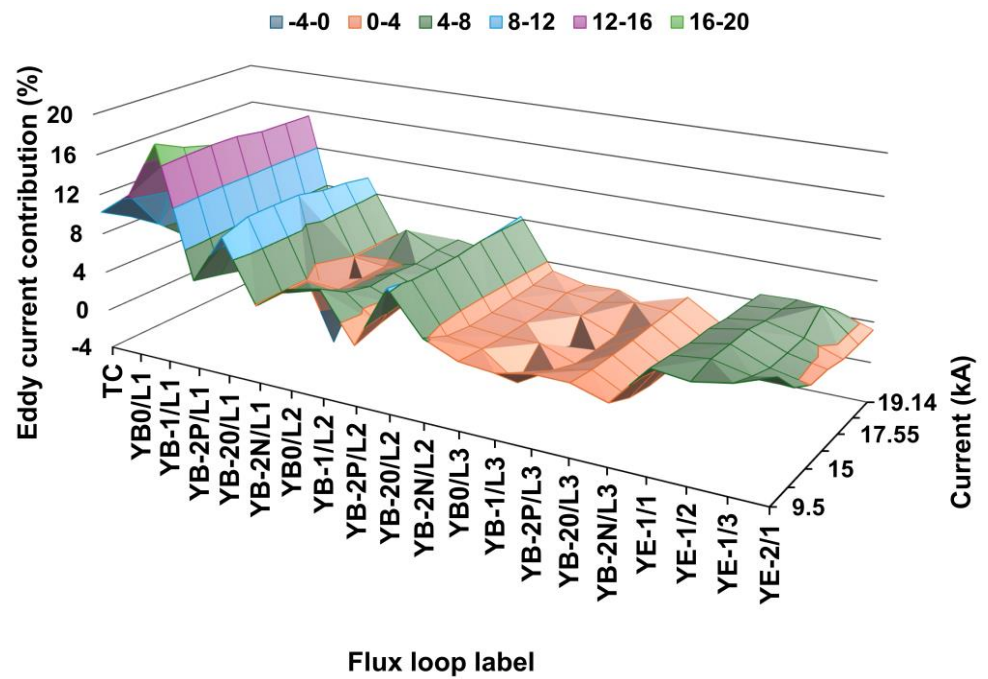

**Figure 7.** The eddy current contributions to 20 flux loops vs. initial currents of the fast discharges.

The flux loops YE−2/2 and YE−2/3 were disconnected when the fast discharges from the currents of 15.221 (on 8 August 2023) and 18.164 (on 22 March 2023) occurred. In the further study, the contributions of the eddy currents into these flux loop signals are taken from the estimations made for the fast discharge from the current of 9.5 kA.

The obtained contributions in each flux loop decrease with increasing initial current values and are in the range from −4.12% (in the flux loop YB−2N/L1 at 15.221 kA) to 18.07% (in the flux loop YB−1/L1 at 9.5 kA). Table 1 presents the values of *E.c. contr* averaged over the flux loops in different yoke layers at each value of solenoid the initial current.

**Table 1.** Eddy current contributions (%) at the magnet steel yoke layers * during the fast discharges.

| Current (kA) | L1 and TC | L2 | L3 | EC−1 | EC−2/1 | Barrel | Endcap | Yoke |
|---:|---:|---:|---:|---:|---:|---:|---:|---:|
| 9.5 | 9.96 ± 5.17 | 5.53 ± 2.43 | 2.67 ± 0.47 | 5.88 ± 0.33 | 7.50 | 6.30 ± 4.53 | 6.28 ± 0.85 | 6.29 ± 4.04 |
| 12.5 | 8.85 ± 4.73 | 4.74 ± 2.02 | 1.66 ± 1.16 | 5.68 ± 0.40 | 5.05 | 5.32 ± 4.29 | 5.52 ± 0.45 | 5.36 ± 3.82 |
| 15 | 8.07 ± 4.74 | 4.37 ± 2.00 | 1.83 ± 0.94 | 5.30 ± 0.73 | 3.77 | 4.97 ± 4.01 | 4.92 ± 0.97 | 4.96 ± 3.58 |
| 15.221 | 6.68 ± 6.81 | 4.38 ± 2.02 | 1.81 ± 0.96 | 5.25 ± 0.75 | 3.70 | 4.44 ± 4.60 | 4.86 ± 0.99 | 4.53 ± 4.11 |
| 17.55 | 7.43 ± 4.83 | 4.48 ± 2.04 | 1.84 ± 0.96 | 4.94 ± 0.46 | 3.19 | 4.76 ± 3.85 | 4.51 ± 0.95 | 4.71 ± 3.45 |
| 18.164 | 7.98 ± 5.09 | 4.86 ± 2.26 | 1.79 ± 1.56 | 5.01 ± 0.25 | 3.16 | 5.27 ± 4.25 | 4.39 ± 1.08 | 5.44 ± 3.80 |
| 19.14 | 7.10 ± 4.89 | 4.69 ± 2.08 | 1.93 ± 0.93 | 4.63 ± 0.15 | 3.20 | 4.73 ± 3.77 | 4.28 ± 0.73 | 4.64 ± 3.37 |

* Only 20 flux loops shown in Figure 7 are used. In the column EC−2/1, the only values for this single loop are presented.

To explain the effect of decreasing the eddy currents with increasing initial current of the CMS solenoid, we can compare the calculated magnetic flux density values in the steel blocks before the fast discharges. Thus, e.g., in the cross-section of the flux loop YB−2P/L2 the calculated axial magnetic flux density is −0.524 T at the current of 9.5 kA, −0.856 T at the current of 15.221 kA, and −1.053 T at the current of 18.164 kA. According to the steel magnetization curve [7] in this block, the values of $dB_z/dI$ in Equation (1) are higher at low current values.

## 3. Results

In Figure 8, the axial $B_z$ (negative) magnetic flux density or vertical $B_y$ (positive) magnetic flux density components calculated in all 22 flux loop cross-sections at the CMS coil currents of 9.5, 12.5, 15, 15.221, 17.55, 18.164, and 19.14 kA are displayed.

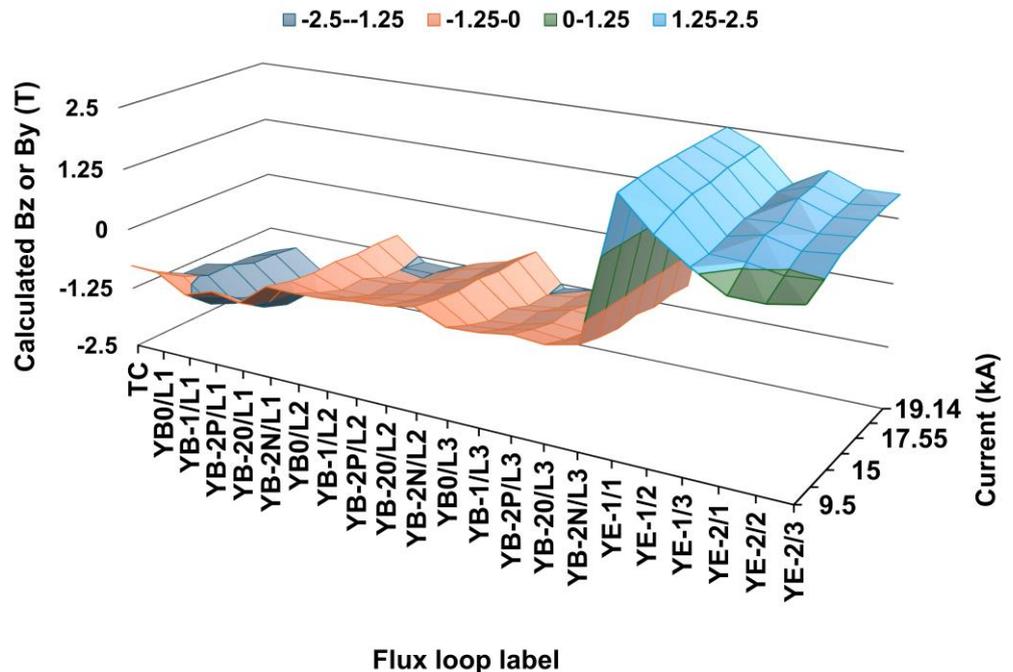

**Figure 8.** Initial axial $B_z$ (negative) magnetic flux density or vertical $B_y$ (positive) magnetic flux density components calculated in all 22 flux loop cross-sections at the seven CMS coil currents.

In the barrel layers, the minimum absolute $B_z$ value of 0.377 T is reached in the cross-section of the flux loop YB−2N/L2 at a current of 9.5 kA. The maximum absolute value of 2.034 T is achieved in the cross-section of the flux loop YB−1/L1 at the current of 19.14 kA. In the endcap disks, the minimum calculated value of 0.947 T is reached in the cross-section of the YE−2/2 flux loop at the current of 9.5 kA. The maximum calculated value of 2.550 T is achieved in the cross-section of the YE−1/1 flux loop at the current of 19.14 kA. The corresponding values measured at the same currents are 0.314 and 2.074 T, and 0.926 and 2.538 T, accordingly. The axial $B_z$ (negative) magnetic flux density or vertical $B_y$ (positive) magnetic flux density measured in all 22 flux loop cross-sections at the CMS coil currents of 9.5, 12.5, 15, 15.221, 17.55, 18.164, and 19.14 kA are presented in Figure 9.

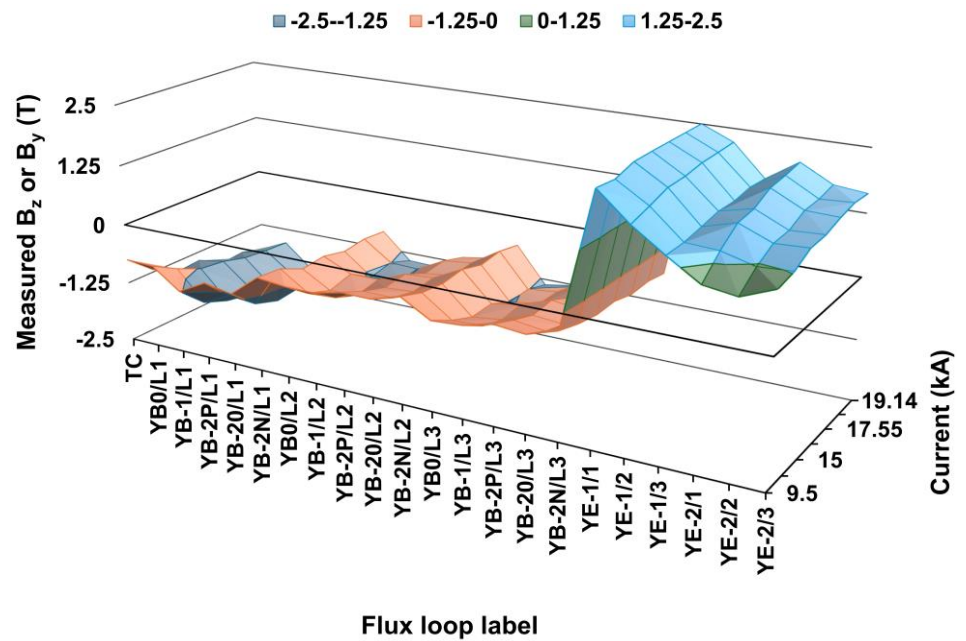

**Figure 9.** Initial axial magnetic flux density or vertical magnetic flux density components measured in all 22 flux loop cross-sections at the seven CMS coil currents.

In Figure 10, the comparisons of measured (*Meas*) and calculated (*Calc*) magnetic flux density components in all 22 flux loop cross-sections at the CMS coil currents of 9.5, 12.5, 15, 15.221, 17.55, 18.164, and 19.14 kA are shown as ratios *(Meas − Calc)/Calc* (%).

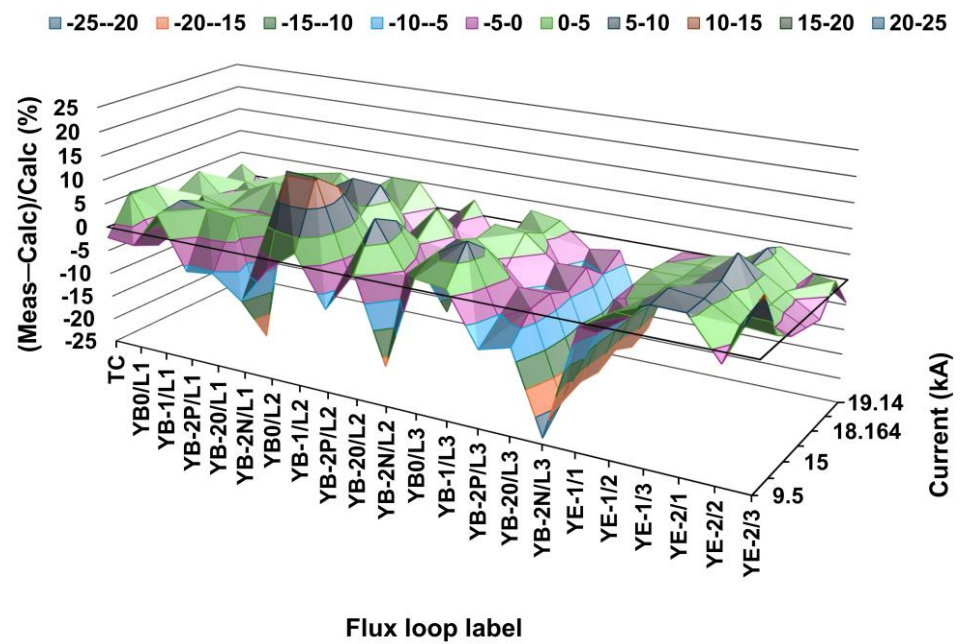

**Figure 10.** Comparison of measured and calculated axial magnetic flux density or vertical magnetic flux density components in all 22 flux loop cross-sections at the seven CMS coil currents.

The ratios described by Equation (1), averaged over the flux loops in different yoke layers at each value of the initial current of the fast discharge, are presented in Table 2.

**Table 2.** The ratios *(Meas − Calc)/Calc* (%) at the magnet yoke layers* for the different initial currents.

| Current (kA) | L1 and TC | L2 | L3 | EC−1 | EC−2 | Barrel | Endcap | Yoke |
|---|---|---|---|---|---|---|---|---|
| **9.5 (new)** | **−3.60 ± 4.91** | **−4.71 ± 7.74** | **−7.39 ± 10.45** | **2.77 ± 3.38** | **4.98 ± 6.87** | **−5.13 ± 7.47** | **3.88 ± 4.99** | **−2.67 ± 7.92** |
| 12.5 (old) | −0.76 ± 10.15 | 7.65 ± 8.51 | −2.85 ± 10.79 | 0.69 ± 5.51 | 3.24 ± 2.22 | 1.22 ± 10.27 | 1.97 ± 4.01 | 1.42 ± 8.91 |
| 15 (old) | 2.30 ± 2.77 | 5.39 ± 7.27 | −2.44 ± 9.45 | 1.59 ± 2.37 | 2.29 ± 3.97 | 1.78 ± 7.13 | 1.94 ± 2.95 | 1.82 ± 6.20 |
| **15.221 (new)** | **0.22 ± 1.30** | **−2.61 ± 6.81** | **−5.73 ± 8.44** | **1.09 ± 1.08** | **1.61 ± 7.57** | **−2.52 ± 6.19** | **1.35 ± 4.84** | **−1.46 ± 6.01** |
| 17.55 (old) | 3.16 ± 2.37 | 3.48 ± 5.53 | −2.47 ± 8.58 | 1.03 ± 1.06 | 1.14 ± 4.15 | 1.50 ± 6.11 | 1.08 ± 2.71 | 1.39 ± 5.33 |
| *18.164 (new)* | *−0.39 ± 0.84* | *−4.58 ± 6.01* | *−6.48 ± 8.17* | *0.09 ± 0.23* | *2.62 ± 3.29* | *−3.60 ± 5.91* | *1.36 ± 2.50* | *−2.25 ± 5.62* |
| 19.14 (old) | 1.66 ± 1.99 | 0.78 ± 5.07 | −3.66 ± 8.21 | 0.14 ± 0.67 | −0.13 ± 4.37 | −0.27 ± 5.64 | 0.005 ± 2.80 | −0.20 ± 4.96 |

* All 22 flux loops shown in Figure 10 are used. The eddy current contributions in the YE−2/2 and YE−2/3 flux loop measurements are taken from the estimations performed at the current of 9.5 kA.

Table 2 contains two sets of comparisons for the new (for the initial currents of 9.5, 15.221, and 18.164 kA) and old (for the initial currents of 12.5, 15, 17.55, and 19.14 kA) flux loop measurements. Comparisons for the new measurements are highlighted in bold as the reference measurements for the eddy current contribution estimations. The line at the current of 18.164 kA, highlighted in italics, exactly corresponds to the comparisons of the measured and calculated magnetic flux density values obtained by averaging the seven sets of measurements performed with the standard CMS magnet ramp-downs [6]. Comparisons for the four old measurements presented in Table 2 are obtained from the set of the fast discharges made in 2006, considering the present eddy current contribution estimations.

From this table, one can observe a systematic difference between the new and old sets of measurements that reflects a different precision of the apparatus used to read out the voltages in 2006 (12-bit DAQ modules) and in 2015 (16-bit DAQ modules). Figure 11 displays the systematic difference in the values of the *(Meas − Calc)/Calc* ratio presented in the last three columns of Table 2.

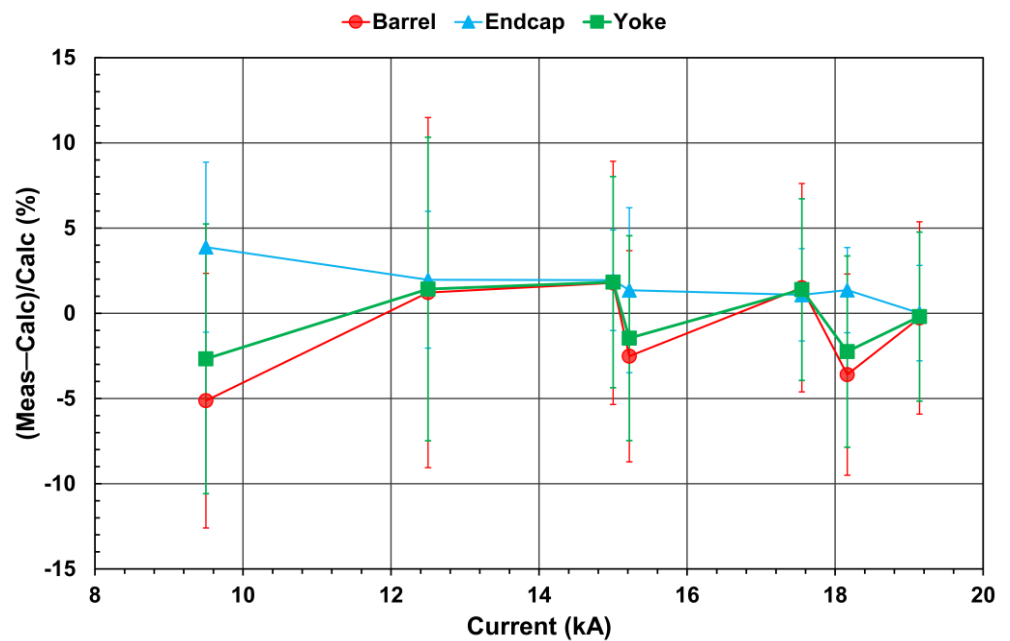

**Figure 11.** Averaged comparison of measured and calculated axial magnetic flux density, $B_z$ (in the 16 barrel flux loops), or vertical magnetic flux density, $B_y$ (in the 6 endcap flux loops), as well as of both components in all 22 flux loop cross-sections (yoke), versus the CMS coil currents of 9.5, 12.5, 15, 15.221, 17.55, 18.164, and 19.14 kA. The currents of 9.5, 15.221, and 18.164 kA are the reference currents, where the eddy current contributions are obtained directly from the data.

In general, the difference between the measured and calculated values of the magnetic flux density in the CMS yoke steel blocks reached several percentage points.

In 2018, the flux loops measurements made earlier in 2006, with the fast discharges from the currents of 17.55 and 19.14 kA, were revised using the eddy current contribution estimations obtained with the fast discharge from the current of 9.5 kA, which occurred in 2017 [14]. The following values of the (*Meas − Calc*)/*Calc* ratio have been obtained with this revision: 0.55 ± 5.98% in the barrel layers and −0.07 ± 1.56% in the endcap disks at the current of 17.55 kA; −1.08 ± 5.52% in the barrel layers and 1.28 ± 1.71 in the endcap disks at the current of 19.14 kA. These values are compatible with the new corresponding values of 1.50 ± 6.11% (barrel layers) and 1.08 ± 2.71% (endcap disks) at 17.55 kA, and −0.27 ± 5.64% (barrel layers) and 0.005 ± 2.80% (endcap disks) at 19.14 kA, which confirms the validity of the new procedure used to estimate the contributions of eddy currents.

## 4. Discussion

To explain the differences observed between the measured and calculated values of the magnetic flux density in the CMS yoke steel blocks, and to find a reason for that, the magnetic flux density *(Meas − Calc)/Calc* ratio in each flux loop cross-section has been considered. Figure 12 shows the comparison of the measured and calculated axial magnetic flux density, $B_z$, or vertical magnetic flux density, $B_y$, in each of the 22 flux loop cross-sections performed in two sets of measurements: the new measurements with the fast discharges from 9.5, 15.221, and 18.164 kA, and the old measurements with the fast discharges from 12.5, 15, 17.55, and 19.14 kA. In this figure, both sets are also combined, and the independent measurements performed in the air gaps between the barrel wheels with the 3D B-sensors [6] are added. The curves in Figure 12 clearly show large discrepancies between the measured and calculated values in the cross-sections of the flux loops near the "negative" (N) edge of the barrel wheel YB−2 in all three layers. This edge is near the large gap between the barrel wheel YB−2 and the endcap disk YE−1, as shown in Figure 1.

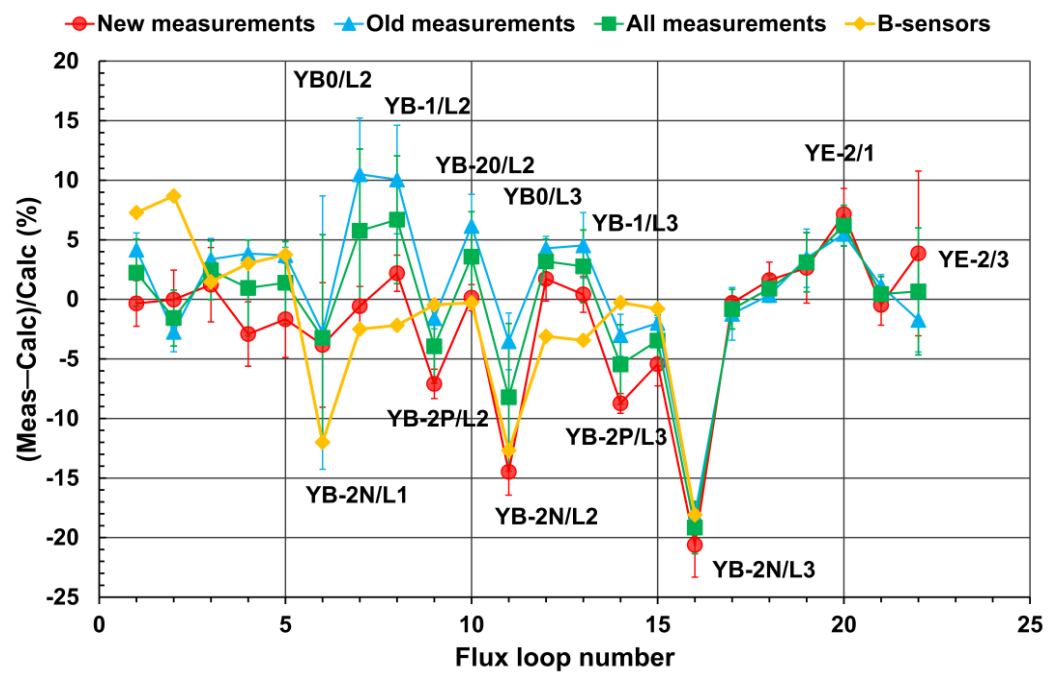

**Figure 12.** Comparison of measured and calculated axial magnetic flux density, $B_z$, or vertical magnetic flux density, $B_y$, in each of the 22 flux loop cross-sections performed in different sets of measurements: new measurements at 9.5, 15.221, and 18.164 kA; old measurements at 12.5, 15, 17.55, and 19.14 kA; and both new and old (all) measurements. The comparisons of the axial magnetic flux density measured with 3D B-sensors [6] and calculated with the CMS magnet model [7] are also displayed.

The 3D B-sensors located at the edges of each layer of the YB−2 wheel also indicate a discrepancy between the calculated and measured axial magnetic flux density. Since the flux loop and B-sensor measurements are independent, the only explanation for these discrepancies is that the CMS magnet model does not describe the magnetic flux well enough in this region. To assess the influence of this effect on the (*Meas* − *Calc*)/*Calc* ratio, the three flux loops YB−2N/L1, YB−2N/L2, and YB−2N/L3 were excluded from consideration. With this approach, the values of the (*Meas* − *Calc*)/*Calc* ratio in the barrel wheel layers became 3.03 ± 3.51% at 17.55 kA and 1.46 ± 2.89% at 19.14 kA. This makes the measurements and calculations compatible within a 3% accuracy. The (*Meas* − *Calc*)/*Calc* ratio is less than 1.5% in 9.1% (the fast discharge from 12.5 kA) to 68.2% (the fast discharge from 18.164 kA) of the flux loop cross-sections, and between 1.5% and 3% in 4.5% (the fast discharge from 18.164 kA) to 36.4% (the fast discharge from 19.14 kA) of the flux loop cross-sections. In contrast, the magnetic flux density distribution in the CMS tracking volume is perfectly homogeneous and is described with the CMS magnet 3D model [7] to within 0.1% accuracy [6].

The practice of 3D modeling and measuring the magnetic field in steel elements of the magnet yoke, as part of a muon detector system, has good prospects for the future. The particle detectors for the Future Circular Electron–Positron Collider (FCC-ee) [16] and the Circular Electron Positron Collider (CEPC) [17] will face a similar problem in 3D modeling and measuring the magnetic field in the detector magnet yoke for the normal operation of the detector's muon spectrometer.

## 5. Conclusions

In this study, the verification of the Compact Muon Solenoid (CMS) magnetic field map in the flux-return magnet yoke steel layers is performed using the magnetic flux density measurements in the 22 cross-sections of the flux loops mounted on 12 yoke steel blocks. Seven fast discharges generated by the CMS coil currents of 9.5, 12.5, 15, 15.221, 17.55, 18.164, and 19.14 kA made it possible to record the flux loop voltages induced by the rapid magnetic flux decay in the flux loop regions and to reconstruct the initial magnetic flux density in the flux loop cross-sections using the off-line integration of these voltages over time. A new method is applied to estimate the contributions to the induced voltages from the eddy currents generated in steel by the fast discharges. The contribution of eddy currents to the flux loop signals reaches 18.07% and decreases to less than 1% for some flux loops, with an increase in the fast discharge initial current. A comparison of the measured and calculated magnetic flux density values in steel shows their compatibility within 3%. In contrast, the magnetic flux density distribution in the CMS tracking volume is described by the three-dimensional model of the CMS magnet with an accuracy of 0.1%.

The practice of 3D modeling and measuring the magnetic field in the steel blocks of the magnet yoke, as part of a muon detector system, has good prospects to be used in the construction and operation of the particle detectors for the Future Circular Electron–Positron Collider and the Circular Electron Positron Collider.

**Author Contributions:** Technical coordination, A.G.; conception, V.K.; engineering integration, B.C., A.G., A.K., M.O., and X.P.; flux loop electronics, A.K., M.O., and X.P.; flux loop read-out software, A.K., M.O., and X.P.; flux loop measurements, B.C., A.K., V.K., M.O., and X.P.; data analysis, V.K.; writing—original draft preparation, V.K.; writing—editing, V.K. and B.C. All authors have read and agreed to the published version of the manuscript.

**Funding:** This research received no external funding.

**Data Availability Statement:** Data are contained within the article.

**Acknowledgments:** The authors are extremely grateful to the CMS Technical Coordinators Alain Hervé, Austin Ball, Wolfram Zeuner, and Paola Tropea of CERN for several years of interest in these studies and for fruitful discussions. We thank Dawn Hudson of CERN for many years' cooperation. We are thankful to Yougang Sun [18] for reviewing the manuscript.

**Conflicts of Interest:** The authors declare no conflicts of interest.